# DEIS: Dependability Engineering Innovation for industrial CPS

E. Armengaud, G. Macher, A. Massoner, S. Frager, R. Adler, D. Schneider, S. Longo, M. Melis, R. Groppo, F. Villa, P. O'Leary, K. Bambury, A. Finnegan, M. Zeller, K. Höfig, Y. Papadopoulos, R. Hawkins, T. Kelly

{eric.armengaud, Georg.macher, alexander.massoner, sebastian.frager}@avl.com, AVL List
{rasmus.adler, daniel.schneider}@iese.fraunhofer.de, Fraunhofer IESE
{simone.longo, massimiliano.melis}@gm.com, General Motors - Global Propulsion System
riccardo.groppo@ideasandmotion.com, Ideas&Motion SRL
federica.villa@polimi.it , Politecnico di Milano
{padraig, kevin}@portablemedicaltechnology.com, Portable Medical Technology Ltd
anita.finnegan@dkit.ie, RSRC at Dundalk Institute of Technology
{marc.zeller, kai.hoefig}@siemens.com, Siemens AG
Y.I.Papadopoulos@hull.ac.uk, University of Hull
{richard.hawkins, tim.kelly}@york.ac.uk, University of York

**Abstract**

The open and cooperative nature of Cyber-Physical-Systems (CPS) poses new challenges in assuring dependability. The DEIS project[1] addresses these challenges by developing technologies that form a science of dependable system integration. In the core of these technologies lies the concept of a Digital Dependability Identity (DDI) of a component or system. DDIs are modular, composable, and executable in the field facilitating (a) efficient synthesis of component and system dependability information over the supply chain and (b) effective evaluation of this information in-the-field for safe and secure composition of highly distributed and autonomous CPS. The paper outlines the DDI concept and opportunities for application in four industrial use cases.

## 1   Introduction

Cyber-Physical Systems (CPS) harbor the potential for vast economic and societal impact in domains such as mobility, home automation and delivery of health.

---

[1] Dependability Engineering Innovation for automotive CPS. This project has received funding from the European Union's Horizon 2020 research and innovation programme under grant agreement No 732242, see www.deis-project.eu

At the same time, if such systems fail they may harm people and lead to temporary collapse of important infrastructures with catastrophic results for industry and society. CPS is the key to unlocking their full potential and enabling industries to develop confidently business models that will nurture their societal uptake. Using currently available approaches, however, it is generally infeasible to assure the dependability of Cyber-Physical Systems. CPS are typically loosely connected and come together as temporary configurations of smaller systems which dissolve and give place to other configurations. The key problem in assessing the dependability of CPS is that the configurations a CPS may assume over its lifetime are unknown and potentially infinite. State-of-the-art dependability analysis techniques are currently applied during design phase and require a priori knowledge of the configurations that provide the basis of the analysis of systems. Such techniques are not directly applicable, can limit runtime flexibility, and cannot scale up to CPS.

The DEIS project addresses these important and unsolved challenges by developing technologies that form a science of dependable system integration. In the core of these technologies lies the concept of a Digital Dependability Identity (DDI [1]) of a component or system. The DDI targets (1) improving the efficiency of generating consistent dependability argumentation over the supply chain during de-sign time, and (2) laying the foundation for runtime certification of ad-hoc networks of embedded-systems. Main contributions of this paper are the introduction of the DDI concept and opportunity analysis for the use of DDI in four relevant industrial use cases from three different domains. The paper is organized as follows: Section 2 introduces the DDI concept, while in Section 3 the four industrial use cases are presented. In Section 4, the opportunities for using the DDI are summarized, and finally Section 5 concludes this paper.

## 2    The Digital Dependability Identity (DDI) concept

In general, a Digital Identity is defined as "the data that uniquely describes a person or a thing and contains information about the subject's relationships"[2]. Applying this idea, a DDI contains all the information that uniquely describes the dependability characteristics of a system or component. This includes attributes that describe the system's or component's dependability behavior, such as fault propagations, as well as requirements on how the component interacts with other entities in a dependable way and the level of trust and assurance, respectively. In general, A DDI is a living model-based modular dependability assurance case. It contains an expression of dependability requirements for the respective component or system, arguments of how these requirements are met, and evidence in the form of safety analysis artifacts that substantiate arguments. A DDI is produced during design, issued when the component is released, and is then continually maintained over the complete lifetime of a component or system. DDIs are used for the integration of components to systems during development as well as for the dynamic integration of systems to "systems of systems" in the field.

A pre-requisite is the availability of a common and machine readable communication language which shall be independent from specific development approaches and tools, and finally enables collaboration between actors in the value chain. Although progress has been achieved with dependability meta-models, e.g. within architecture description languages like EAST-ADL and AADL, there is still a lack of a common model for the communication of dependability information. The recently released Structured Assurance Case Metamodel (SACM) [3] defines a meta-model for representing structured assurance cases, i.e. a set of auditable claims, arguments, and evidence created to support the claim that a defined system/service will satisfy the particular requirements. The SACM will be the meta model for the externally visible DDI interface, whereas the internal "logic" will be described by other, already existing approaches.

Managing the variability of a system is a key challenge in the industrial context, e.g., during different development iterations or for the management of product variants. This requires that DDIs support change impact analyses that enable a prediction about whether a component will fit into multiple product variants, to help reduce re-assurance effort. As these models would be modularized in the level of components, DDI would allow to conduct such analyses across different DDI employing different techniques (such as C2FTs or Hip-HOPS [5] [6]).

In connected CPS, dependability cannot be fully assured prior to deployment, because systems will dynamically interconnect (with $3^{rd}$ party systems) and form systems of systems with largely unpredictable consequences for dependability. In order to assure the dependability of such in-field integrations we propose automated DDI-based dependability checks at integration time. DDIs therefore must become executable specifications accompanying systems through their complete lifecycle, not simply digital artifacts that cease in their utility after deployment of the system.

A central question is, however, to which extent "safety intelligence" can be shifted from development time into runtime (see Fig 1). A possible first step is to use runtime certificates such as ConSerts [4]. In this case, the assurance case is still completely designed and managed at development time, only series of unknown context-dependencies are formalized and shifted into runtime, "known unknowns". The next step would be to deal with "unknown unknowns" at the design level, with the requirements and thus the safety goals still being determined by human engineers. While SafetyCases@Runtime would be capable to run predefined validation and verification activities at runtime to obtain required evidences, V&V-Models@Runtime additionally support the modification of V&V models, e.g., the modification of test cases or pass/fail criteria. As soon as the adaptation to 'unknown unknowns' also requires an adaptation of requirements, it is additionally necessary to adapt the hazard and risk analysis and the resulting safety goals at runtime. With that last step, a crosscutting aspect of the whole dependability lifecycle would be shifted into runtime, which is today hardly conceivable.

To achieve this vision, in DEIS we develop DDIs as modular, composable, and executable model-based assurance cases with interfaces expressed in SACM and internal logic that may include fault trees, state automata, Bayesian networks and fuzzy models for representing uncertainty. Modularity means DDIs apply to units at different levels of design, including the system itself, its subsystems and components. By being composable, the DDI of a unit can be derived in part from its constituent elements. To simplify DDI construction and synthesis, in DEIS we experiment with techniques for automatic, model-based construction of DDIs via automatic allocation of safety requirements and auto-generated model-based dependability analyses [5]. To enable executability, we explore the use of the models that represent the internal logic of DDIs for dynamic detection and prognosis of risks, considering among other factors limitations in observability and environmental uncertainties [4][6]. Within the scope of DDI modelling, we include examination of implications of security breaches on safety.

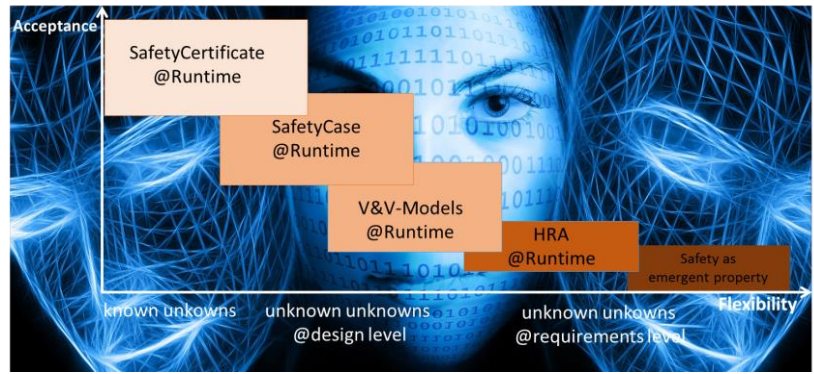

**Fig. 1.** The different levels of safety variability

## 3  The four industrial use cases in DEIS project

### 3.1 Automotive: development of a stand-alone system for intelligent physiological parameter monitoring

Global healthcare is worth 1.5$ trillion, and a part of that, around 11$ billion, is focusing on physiological portable monitor system (e.g. wearable and quantified self) [8], [9]. Moreover, a recent New AAA Foundation report reveals drivers spend an average of more than 17,600 minutes behind the wheel each year[2]. Future frontiers of smart mobility rise the need to measure physiological parameter of the driver and passengers and to evaluate the health of the occupants. This to enable autonomous driving features according to driver health state, and enable potential B2B and B2C opportunities.

DDIs are a revolutionary concept for the automotive domain and can open up new prospects, providing means for ensuring dependability among CPS and improving safety of driver and passengers, taking actions (supporting with autonomous drive and/or health services) in case of emergency/needs. The aim of this use case is the introduction of a dependable physiological monitoring system applied to a connected vehicle, capable to identify physiological parameters and to evaluate the health of driver and other occupants. The proposed environment (see Fig 2) contains a comprehensive package of technologies, tools and services that support the drive session in evaluating driver health state, and enable autonomous driving features according to the needs. The same acquired parameters are also transmitted to a cloud-based system for online support and health analysis which is provided back to the customer as additional service.

Whilst there is an opportunity to improve safety, at the same time the system is potentially prone to security attack. One of the challenges will be to maximize the *Confidentiality* of health data transmitted, and guarantee *Safety* and *Plausibility* analysis on acquired signals, as well as *Integrity* on communication among different CPSs.

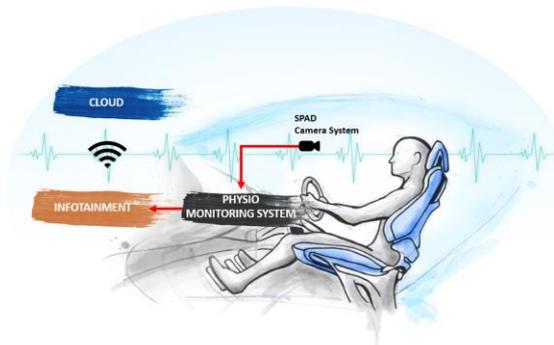

**Fig. 2.**   Intelligent physiological parameter monitoring – overview

Application of DDIs shall help protection of sensitive information sent through the network, granting the authenticity on actors involved, senders and receivers. Crypto methodologies should be used to avoid tampering and prevent corruption during transmission. The DDI shall also help to guarantee a certain level of privacy on the sensitive physiological data collected and transmitted. The Cloud-based system shall collect all data of each CPS and keep data in synch for each session. Overall, the system shall be capable to access all the services and send/receive DDIs in any operative and network conditions. The physiological monitor system shall be representative of the true driver health condition and noise resistant. This is because, depending on the driver health condition, the system should react promptly activating a predefined action, minimizing delays or applying workarounds to possible network unavailability. The physical system should remain compliant with the automotive standard (ISO 26262) in the case that autonomous drive action is enabled.

## 3.2 Automotive: enhancement of an advanced driver simulator for evaluation of automated driving functions

This use case introduces a novel driver simulator based on AVL VSM[3] for the comprehensive evaluation of complex autonomous driving functions. The proposed environment contains an extensive package of tools and services that support the OEM in the prediction of vehicle behavior, and enables improvement of various vehicle attributes from the initial concept to the testing phase.

An approach for the optimal control of a fully electric vehicle and its powertrain approaching a road segment with Multiple Traffic Lights (TL) has been presented in [10]. A system referred to as the Traffic Light Assistant (TLA) was developed to in order to take over longitudinal control of the vehicle, optimizing the velocity trajectory when approaching multiple traffic lights in traffic. The main aims of the system are to reduce energy consumption and $CO_2$ emission, reduce the number of vehicle stops and waiting times, and introduce smoother speed profiles (see Fig. 3). The presented TLA approach has since been further developed to work with other powertrain topologies [11] [12] and was demonstrated in real-time for conventional vehicles on the powertrain testbed within the FFG Austrian Funded R&D Project TASTE (Traffic Assistant Simulation and Testing Environment).

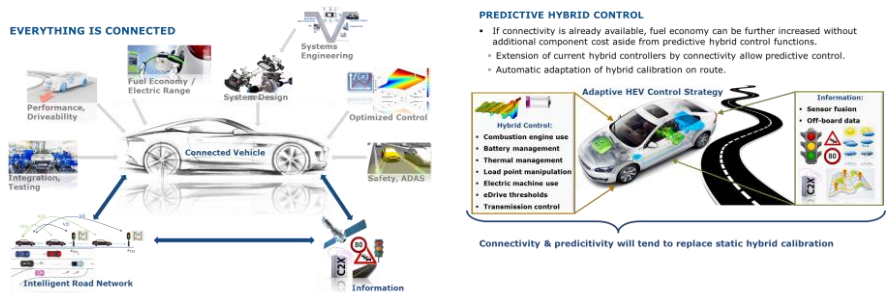

**Fig. 3.** The connected powertrain and autonomous driving functions allow increased energy efficiency (reduction of fuel consumption, emissions)

Previous work has focused on controlling an "ego vehicle" equipped with TLA, assuming complete knowledge about the road conditions and traffic light signal phasing. In reality, not all information may be accessible at all times to every vehicle. For example, if the current state of a traffic light or a pedestrian at the side of a crossing is observed by an on-board camera or local V2I communication, the information would only be available to a vehicle close by. Such a vehicle, if equipped with automated driving functions, would adapt its velocity accordingly to stop in front of the traffic light, or to safely let the pedestrian pass. However, a following vehicle might not directly see the traffic light or the pedestrian and therefore could not anticipate the behavior of the preceding vehicle. In this case, a reaction can only occur based on observation of the behavior of the preceding vehicle. It is possible to improve energy efficiency, if the state of traffic light or the presence of pedestrian are known in advance.

---
[3] https://www.avl.com/-/avl-vsm-vehicle-simulation

Once more the opportunity for increased efficiency and safety comes with increased security threats. Thus, DDIs will explore how to improve control strategies to increase energy efficiency while ensuring safety at the same time. Equally important would be to analyze and evaluate the impact of possible security attacks, and further define strategies to avoid these attacks or mitigate their effects [13].

### 3.3 Railway: Enabling Plug-and-play scenarios for heterogeneous railway systems

With a total market volume of 47 billion € and nearly 400,000 employees in the European rail industry (by 2014), the rail industry plays a key role in mobility in the EU [14]. The European railway domain has to cope with the challenging situation of heterogeneous systems of systems with different standards and system qualities, e.g. interoperability between train side and track side systems. Moreover, railway systems typically comprise both legacy systems and systems at the cutting edge of technology, which are potentially not designed to be integrated but need to interact in operation.

The European Commission's long-term objective is to achieve a *Single European Railway Area* to deliver the benefits of market opening and interoperability as well as to reduce the life-cycle costs (i.e. the costs of developing, building, maintaining, operating, renewing and dismantling) of rolling stock and on-board signalling systems (50% reduction by 2030) [15]. The challenges to reach this goal not only include harmonizing technical interfaces, but also a common signalling system.

The European Train Control System (ETCS) provides standardized train control in Europe and eases travelling with trains crossing the borders of all countries in Europe. Due to historical reasons, different trains and also different track side solutions (such as balises, track circuits or GSM-R transmission) are often present even within one country. To overcome the challenges, the EC fosters the technical harmonization within the European railway sector ensuring interoperability in the railway domain. Thus, enabling "plug & play" of railway systems scenarios as a long-term objective [16]. However, in such "plug & play" scenarios guaranteeing system dependability requirements pose new challenges. Consequently, certification activities (w.r.t. CENELEC standards EN 50126 & 50129) require accurate planning and must react quickly to changes within the system development process. Systems of systems in the railway domain are also produced by various stakeholders in the value chain (such as national or even regional public transport authorities, national safety authorities, railway undertaking, OEMs, suppliers, etc.)

and, therefore, safety information about components and subsystems (rolling stock, track-side and railway systems) need to be interoperable and exchangeable.

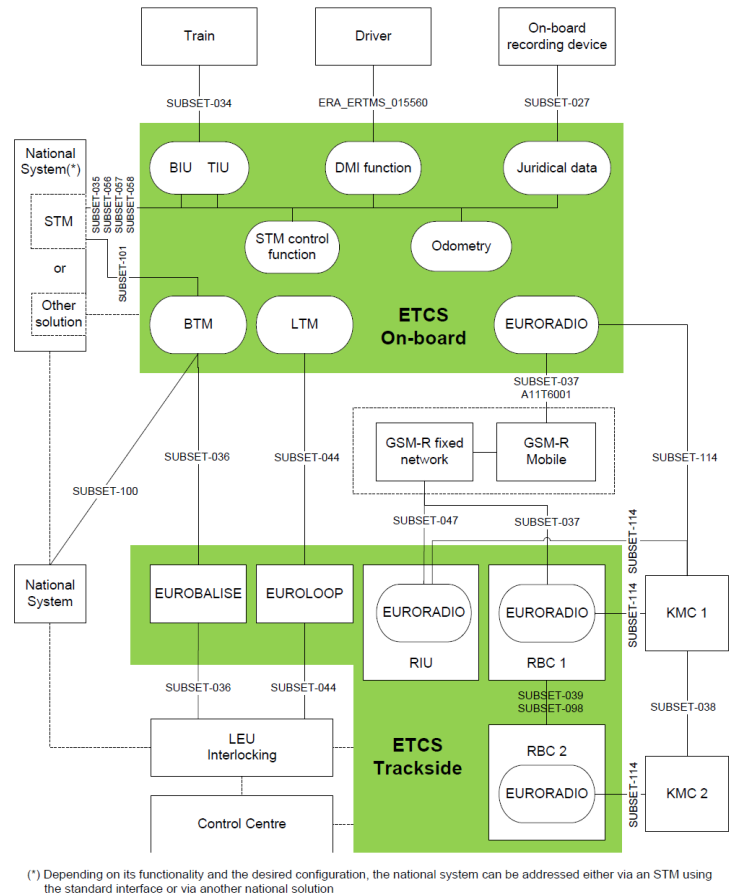

**Fig. 4.** ERTMS/ETCS Reference Architecture

As illustrated in Fig 4, ETCS consists of an on-board and a trackside system. Both sub-systems must fulfill the safety requirement as defined in the ERTMS/ETCS specification. Thereby, specific hazards a`s well as tolerable hazard rates are apportioned to each system. Moreover, trackside and on-board system are often provided by different vendors. ETCS on-board and trackside systems must meet the specified safety requirements specified in the specification in order to be safely integrated in any interoperable railway system.

In this use case, we use DDIs to interchange safety-relevant information (including e.g. safety requirements, models, and assessments) during the development life-cycle of the trackside and the on-board ETCS units. We thereby target to demonstrate the improvement of interoperability in the area of safety engineering across companies, railway operators as well as safety authorities. Hence, we show how time and effort in the certification of systems (or sub-systems) can be reduced significantly by interchanging and reusing dependability information across the value chain of the railway domain.

### 3.4 Healthcare: enhancement of clinical decision app for oncology professional

The healthcare industry globally is moving towards Electronic Health Records (EHR's) for the management and storage of patient data[4]. ONCOassist[5] is a clinical decision support app for oncology professionals, see Fig 5. It contains all the key oncology decision support tools oncology professionals need and makes them available in an easy to access and interactive format at point of care. It is CE approved, meaning it is classified as a medical device and was developed in a regulated environment using ISO13485 and IEC62304.

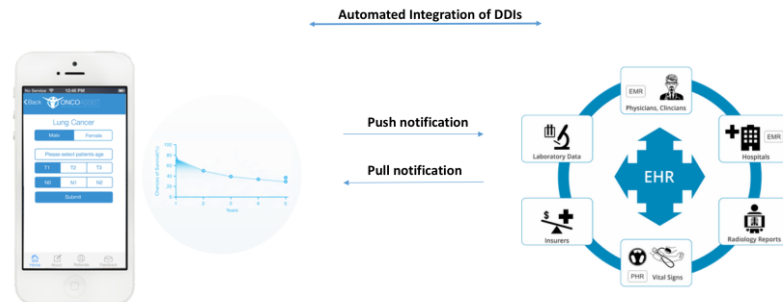

**Fig. 5.** ONCOAssist - Clinical decision support app for oncology professionals

The demand for clinical decision support apps is growing rapidly because of (a) genomic sequencing, (b) aging populations, and (c) new targeted therapies. These changes in the market mean the decision making process for oncology professionals is becoming more difficult and complex and they need systems to support them with this. ONCOassist is addressing this market need by putting key decision support information and tools in the palm of the hand of the oncology professionals. The ability to transfer new dependability-relevant data from service provider to a data repository is increasingly important. This new dependability-relevant information could be integrated in an existing data record within the data repository in a way that preserves the intended meaning of the information being transferred. The service provides dependability information in form of DDIs and exchanges them with the data repository. When the systems meet at runtime, they use the DDI's to check if they can provide together new services in a dependable way. If this check is positive, then they start collaborating and providing the service, exchanging data and information. This in turn means that we can provide new services with confidence on the integrity of information handled. DDIs in this context can result in ad-hoc (in the wild) integration of ONCOassist with EHRs and hospital information systems, in turn leading to context aware decision support for clinicians.

---

[4] http://www.healthcareitnews.com/news/precision-medicine-growth-hinges-electronic-health-records
[5] http://oncoassist.com/

## 4      Opportunities for DDI applications

As an outcome of the analysis of the use cases presented in section 3, four scenarios for the usage of DDI in industrial application were identified

**Sc1: Creation of integrated system assurance case for distributed development of dependable systems**

In a supply chain, the OEM provides system dependability requirements which must be taken into account by suppliers. The component supplier engineers create an assurance argument for their component including the definition of assurance claims for the component, evidence relating to those claims and argument explaining how the evidence supports the claims. Information and evidence from the assurance case is translated into an SACM model for exchange as a DDI. This activity may be automated for component and system models used in the assurance case. Once the translation to DDIs is complete the higher-level assurance case for the system is created by the OEM via analysing and integrating information exchanged in the DDIs. This activity may be at least partially automated in order to support dynamic assurance case creation at run-time. Expected result is a strong efficiency increase by compiling and updating complex system assurance cases thanks to the availability of the SACM standard, the DDI framework and the related tools that automate part of the task.

**Sc2: Runtime monitoring and optimization with respect to dependability**

In an open system of systems, loosely connected systems collaborate to provide dependable services to users. Service are enabled by the collaboration and adapted at runtime to fulfill certain dependability requirements. When systems meet at runtime, the DDIs (set-up during development time) are used to check if they can provide together new services in a dependable way. If this check is passed, then collaboration can occur. To assure and maintain dependability properties, DDI are continually executed to adapt the collaboration accordingly. Expected results shall impact the deployment and adaptation of complex systems of systems to maximize functionalities and performances, while securing dependability.

**Sc3: Next Generation Connected Dependable Functionalities**

In the next generation of connected vehicles, infrastructure- and 3[rd]-party software providers will be delivering situation-dependent services or additional ser-vices like apps for smart phones. In this context, DDIs will support the dependability of dynamic implementations which are based on service oriented architectures, thus enabling novel situation dependent features for example in the context of autonomous and automated driving

**Sc4: Dependable runtime integration for exchange of information**

In this scenario, DDIs oversee the transfer of dependability-relevant data from a service provider to a data repository in an open system where the connection between provider and repository is dynamically redefined. Data integrity and security are the prime concerns addressed here.

## 5      Conclusions

The physical and digital worlds are currently merging, leading to a largely connected globe. However, developments like IoT and CPS pose enormous ethical, economic and related technical questions which we should address responsibly in the traditions of scientific method. A key challenge that we identified and discussed here is that it is currently impossible to assure the dependability of smart CPS such as autonomous cars, swarms of drones, or networks of telehealth devices. Such systems are loosely connected and come together as temporary configurations of smaller systems which dissolve and give place to other configurations. The configurations a CPS may assume are unknown and potentially infinite. In addition, as systems connect, emergent system behaviours may arise in ways that are difficult to predict from simple superposition of the behaviour of individual system elements. State-of-the-art dependability analysis techniques are currently applied during the design phase and require full a priori knowledge of system configurations. Such techniques are not directly applicable to open and dynamically reconfigured CPS.

Moving beyond the challenge, we introduced the DEIS H2020 research project, an effort focusing on the problems discussed above. The project is developing a foundation of methods and tools that lays the groundwork for assuring the dependability of CPS. In the core of this work lies the novel concept of a DDI for components and systems. DDIs are planned as an evolution of current modular dependability specifications and model aspects of the safety, reliability and security "identity" of the component. They are produced during the design phase and their profiles are stored in the "cloud" to enable checks by third parties. They are composable and executable, and facilitate dependable integration of systems into "systems of systems". The paper introduced early work in this area and discussed a wide range of cross-sectoral opportunities for industrial application. We hope to be able to report soon providing reflections on case studies and results of evaluations.

# 6 References

[1] Schneider, D., Trapp, M., Papadopoulos, Y., Armengaud, E., Zeller, M., & Höfig, K. (2015). WAP: Digital dependability identities. *26th International Symposium on Software Reliability Engineering (ISSRE'15)*, (pp. 324-329)

[2] Windley, P., "Digital Identity," O'Reilly Media, 2005

[3] Structured Assurance Case Metamodel (SACM), version 2.0, Object Managment Group (OMG), 2016, available at http://www.omg.org/spec/SACM/2.0/Beta1/PDF/

[4] Schneider, D., & Trapp, M. (2013). Conditional Safety Certification of Open Adaptive Systems. ACM Trans. Auton. Adapt. Syst. 8, 2, Article 8 (July 2013), (p. 20).

[5] Azevedo, L.; Parker, D.; Walker, M.; Papadopoulos, Y.; Esteves Araujo, R.: Assisted Assignment of Automotive Safety Requirements. IEEE Software, 31(1), pp. 62-68, 2014

[6] Dheedan A., Papadopoulos Y., Multi-Agent Safety Monitoring System, IFAC Proceedings, 43(4):84-89, 2010

[7] Luckham, D.: The Power of Events: an Introduction to Complex Event Processing in Distributed Enterprise Systems. Addison-Wesley, 2001

[8] HRI analysis and Centers for Medicare and Medicaid Services National Health Expenditures (2012)

[9] PwC Health Research Institute, New Health entrants 2015 - Global health's new entrants: Meeting the world's consumer, 2015, available at https://www.pwc.com/mx/es/industrias/archivo/2015-02-global-healthcare-new-entrants.pdf

[10] Ferreira-Parrilla, A., et. al.: Traffic Light Assistant System for Optimized Energy Consumption in an Electric Vehicle, ICCVE, 2014.

[11] Jones, S. et.al., V2X Based Traffic Light Assistant for Increased Efficiency of Hybrid & Electric Vehicles, Automotive meets Electronics Congress, Conference Paper, 2016.

[12] Jones, S. et.al., V2X Based Traffic Light Assistant for Increased Efficiency of Hybrid & Electric Vehicles, VDI Wissensforum, Conference Paper, 2016.

[13] Armengaud, E., Höller, A., Kreiner, C., Macher, G., & Sporer, H. (2015). A Combined Safety-Hazards and Security-Threat Analysis Method for Automotive Systems. SAFECOMP international conference on computer safety, reliability and security.

[14] UNIFE. (2014). European Rail Industry Guide, available at http://www.unife.org/index.php?option=com_attachments&task=download&id=499

[15] EC. (2011). Roadmap to a Single European Transport Area - Towards a competitive and resource efficient transport system. European Commission.

[16] Shift2Rail. (2015). *Shift2Rail Strategic Master Plan.* Shift2Rail., available at https://ec.europa.eu/transport/sites/transport/files/modes/rail/doc/2015-03-31-decisionn4-2015-adoption-s2r-masterplan.pdf

# 7 Full Authors' Information

Eric Armengaud, Georg Macher, Alexander Massoner, Sebastian Frager
AVL List GmbH,
Hans List Platz 1, 8020 Graz, Austria
{eric.armengaud, georg.macher , alexander.massoner, sebastian.frager}@avl.com

Daniel Schneider, Rasmus Adler
Fraunhofer IESE,
Fraunhofer Platz 1, 67663 Kaiserslautern, Germany
{daniel.schneider, rasmus.adler}@iese.fhg.de

Simone Longo, Massimiliano Melis
General Motors - Global Propulsion System
Corso Castelfidardo, 36, 10138 Torino, Italy
{simone.longo, massimiliano.melis}@gm.com

Riccardo Groppo
Ideas & Motion SRL
Via Moglia 19, 12062 Cherasco (CN), Italy
riccardo.groppo@ideasandmotion.com

Federica Villa
Politecnico di Milano
Piazza Leonardo da Vinci, 32, 20133 Milano, Italy
federica.villa@polimi.it

Padraig O'Leary, Kevin
Portable Medical Technology Ltd
Google Campus, 4-5 Bonhill St, London EC2A 4BX, United Kingdom
{padraig, kevin}@portablemedicaltechnology.com

Anita Finnegan
RSRC at Dundalk Institute of Technology
Marshes Upper, Dundalk, Co. Louth, Irland
anita.finnegan@dkit.ie

Marc Zeller, Kai Höfig
Siemens AG, Corporate Technology
Otto-Hahn-Ring 6, 81739 Munich, Germany
{marc.zeller, kai.hoefig}@siemens.com

Yiannis Papadopoulos
University of Hull
Cottingham Rd, Hull HU6 7RX, United Kingdom
Y.I.Papadopoulos@hull.ac.uk

Richard Hawkins, Tim Kelly
University of York
Heslington, York YO10 5DD, United Kingdom
{richard.hawkins, tim.kelly}@york.ac.uk

# 8 Keywords

Safety, security, Digital Dependability Identity (DDI), automotive, railways, healthcare